\documentclass[showpacs,showkeys,eqsecnum,prd,aps,twoside,twocolumn]{revtex4-1}
\usepackage{amsfonts,amsmath,amssymb,bm,graphicx}

\begin{document}


\title{Lepton asymmetry growth in the symmetric phase of an electroweak plasma with hypermagnetic fields versus its washing out by sphalerons}

\author{Maxim Dvornikov${}^{a,b}$}
\email{maxim.dvornikov@usp.br}
\author{Victor B. Semikoz${}^{a}$}
\email{semikoz@yandex.ru}
\affiliation{${}^{a}$Pushkov Institute of Terrestrial Magnetism, Ionosphere
and Radiowave Propagation (IZMIRAN),
142190 Troitsk, Moscow Region, Russia
\\
${}^{b}$Institute of Physics, University of S\~{a}o Paulo, CP 66318, CEP 05315-970 S\~{a}o Paulo, SP, Brazil;\\
}

\date{\today}

\begin{abstract}
We study lepton asymmetry evolution in plasma of the early Universe before the electroweak phase transition (EWPT)
accounting for chirality flip processes via Higgs decays (inverse decays) entering equilibrium at temperatures
below $T_\mathrm{RL} \simeq 10\ \mathrm{TeV}$, $T_\mathrm{EW} < T < T_\mathrm{RL}$. We solve appropriate kinetic equations for leptons and Higgs bosons taking into account the lepton number violation due to Abelian anomalies for right and left electrons  and neutrinos in the self-consistent hypercharge field obeying Maxwell equations modified by the contribution of the Standard Model of electroweak interactions. The violation of left lepton numbers and the corresponding violation of the baryon number due to sphaleron processes in symmetric phase is taken into account as well. Assuming the Chern-Simons wave
configuration of the seed hypercharge field, we get the estimates of baryon and lepton asymmetries evolved from the primordial right electron asymmetry existing alone as partial asymmetry at $T \geq T_\mathrm{RL}$. One finds a strong dependence of the asymmetries on the Chern-Simons wave number. We predict a nonzero chiral asymmetry $\Delta \mu=\mu_{e_\mathrm{R}} - \mu_{e_\mathrm{L}}\neq 0$ in this scenario evolved down to the EWPT moment that can be used as an initial value for the Maxwellian field evolution after EWPT.
\end{abstract}



\maketitle

\section{Introduction\label{INTR}}
The nature of the initial fields that seed subsequent dynamo for observed galactic magnetic fields is largely unknown
\cite{Kulsrud:2007an,Kronberg:1993vk}. It might be that seed fields are produced during the epoch of the galaxy formation, or ejected by first supernovae or active galactic nuclei. Alternatively to this astrophysical scenario the seed fields might originate from much earlier epochs of the Universe expansion, down to the cosmological inflation epoch~\cite{Grasso:2000wj}. There are first observational indications of the presence of cosmological magnetic fields (CMF) in the intergalactic medium which may survive even till the present epoch~\cite{Neronov:1900zz,Neronov:2009gh}.

It is well known that Maxwellian CMF might arise during the electroweak phase transition (EWPT) from massless (long-range) hypercharge fields $Y_{\mu}$ existing in primordial plasma before EWPT~\cite{Giovannini:1997eg}. Note that long-range non-Abelian magnetic fields (corresponding to, e.g., the color SU(3) or weak SU(2) groups) cannot exist because at high temperatures the non-Abelian interactions induce a ``magnetic" mass gap $\sim g^2T$.

The Faraday equation which governs evolution of magnetic (hypermagnetic) fields in the early Universe  depends crucially on the helicity parameter $\alpha$ which is a {\it scalar} owing to the {\it parity violation} in the Standard Model (SM) and having different forms before and after EWPT. We remind that the standard magnetohydrodynamic (MHD) parameter, which is generated by vortices in plasma $\alpha_\mathrm{MHD} \sim \langle {\bf v}\cdot (\nabla\times {\bf v}) \rangle$ is \textit{pseudoscalar} according to the parity conservation in QED plasma. Obviously in the isotropic early Universe such vortices are absent, at least at large scales we consider here.

Before EWPT in the symmetric phase of primordial plasma the helicity parameter $\alpha_\mathrm{Y}$ results from the Chern-Simons (CS) anomaly term in the SM Lagrangian~\cite{Giovannini:1997eg,Redlich:1984md} $L_\mathrm{CS}= (g'^{2}\mu_{e_\mathrm{R}}/4\pi^2)(\mathbf{B}_\mathrm{Y} \cdot {\bf Y})$, which violates parity. Here, $g' = e/\cos \theta_\mathrm{W}$ is the $U_\mathrm{Y}(1)$ gauge coupling, $\mathbf{B}_\mathrm{Y}=\nabla\times {\bf Y}$ is the hypermagnetic field, and $\mu_{e_\mathrm{R}}$ is the chemical potential for right-handed electrons (positrons). The {\it polarization origin} of the CS term $L_\mathrm{CS}$ was elucidated in Ref.~\cite{Semikoz:2011tm} as an effect of comoving right electrons and positrons along $\mathbf{B}_\mathrm{Y}$ which have opposite spin projections on $\mathbf{B}_\mathrm{Y}$ and populate the main Landau level with slightly different densities due to $\mu_{e_\mathrm{R}}\neq 0$.

While such CS term vanishes in broken phase at $T<T_\mathrm{EW}$~\cite{Laine:1999zi}, there appears a similar polarization
effect~\cite{Semikoz:2003qt} due to weak interaction of neutrinos (antineutrinos) with polarized electrons (positrons) given by the axial vector force ${\bf F}^{(A)}_{\sigma}=-\nabla V^{(A)}_{\sigma}$ arising from the parity violating part of weak interactions between neutrinos of the flavor $a=e,\mu,\tau$ and charged leptons,
$$
  V_{\sigma}^{(A)}(\mathbf{x},t) = G_\mathrm{F}c^{(A)}_a(\mathbf{M}^{(\sigma)}\cdot \delta \mathbf{j}^{(\nu_a)}(\mathbf{x},t)).
$$
This axial vector force acts only on those polarized electrons and positrons which contribute to the partial magnetization $\mathbf{M}^{(\sigma)}=\mu_\mathrm{B}\ {\rm sign}(\sigma)\ n_{0\sigma}\mathbf{B}/B$ with the corresponding densities $n_{0\sigma}$ of electrons ($\sigma=-$) and positrons ($\sigma=+$) at the main Landau level. Here $G_\mathrm{F}$ is the Fermi constant, $c_a^{(A)}=\mp 0.5$ is the axial coupling in the Weinberg-Salam model, the upper (lower) sign stays for electron (muon, tau) neutrinos, and $\mu_\mathrm{B}$ is the Bohr magneton. Note that the neutrino current density asymmetry $\delta \mathbf{j}^{(\nu_a)}=\mathbf{j}_{\nu_a} - \mathbf{j}_{\bar{\nu}_a}$ is the polar vector and the magnetization $\mathbf{M}^{(\sigma)}$ is the axial vector. Thus, the weak axial-vector force separates electric charges causing the relative drift velocity and the corresponding electric current that leads to the generation of an additional electromagnetic field component resulting from weak interactions $\mathbf{E}_\mathrm{weak}\sim \alpha \mathbf{B}\sim G_\mathrm{F}$. Unfortunately the $\alpha$-helicity parameter produced by this mechanism is very small. Moreover, it depends on such additional parameter as a neutrino gas inhomogeneity scale $\lambda_\mathrm{fluid}^{(\nu)}$, $\alpha\sim G_\mathrm{F}(T/\lambda_\mathrm{fluid}^{(\nu)})(\delta n_{\nu}/n_{\nu})$~\cite{Semikoz:2003qt}. It should be mentioned that the new opportunity for the Maxwellian CMF generation found in recent works~\cite{Boyarsky:2011uy,Boyarsky:2012ex} seems to be very intriguing.

In Ref.~\cite{Boyarsky:2011uy} it was shown that the evolution of Maxwellian magnetic fields in a primordial plasma at temperatures $T\geq 10\ \mathrm{MeV}$ is strongly affected by the quantum chiral anomaly proportional to the difference of right-handed and left-handed electron chemical potentials $\Delta \mu(t)=\mu_{e_\mathrm{R}} - \mu_{e_\mathrm{L}}$ evolving in a self-consistent way. Such a difference defines the new helicity parameter $\alpha_\mathrm{new}(t)=\alpha_\mathrm{em} \Delta \mu (t)/\pi\sigma_\mathrm{cond}$ in the modified Faraday equation that governs evolution of Maxwellian fields just after EWPT at temperatures $10\ \mathrm{MeV}<T<T_\mathrm{EW}$ (see comments below in Appendix~\ref{ORIGIN}). Here $\alpha_\mathrm{em} = 1/137$ is the fine-structure constant and $\sigma_\mathrm{cond}$ is the plasma conductivity.

The goal of the present work is a more careful analysis of the {\it initial chiral anomaly parameter} $\Delta \mu (T_\mathrm{EW})$ arising in hypermagnetic fields before EWPT and accounting for the Higgs (inverse) decays which change chiralities of right- and left-handed electrons (positrons). While the sphaleron processes are always switched off in the Higgs phase, in the symmetric phase at $T > T_\mathrm{EW}$, conversely, we should take them into account since lepton and baryon numbers are violated even in the absence of hypermagnetic fields. The plan of our paper is the following. In Sec.~\ref{EQUILIBRIUM} we formulate equilibrium conditions in the symmetric phase of primordial plasma and comment on the corresponding set of chemical potentials.  In the main Sec.~\ref{KINETICS} we derive kinetic equations for appropriate asymmetries of right- and left-handed electrons (positrons) in the presence of Abelian anomalies for fermions, as well as the kinetic equation for the Higgs boson asymmetry accounting for both inverse and direct Higgs decays. Such an extension of the kinetic approach, considered earlier in our paper~\cite{Dvornikov:2011ey}, allows us to correct the value  $\Delta \mu (T_\mathrm{EW})$ accounting for the Higgs boson asymmetry evolution.
In Sec.~\ref{CONSERVLAWS} we check conservation laws and calculate the baryon asymmetry of the Universe (BAU) through leptogenesis in hypermagnetic fields. In Sec.~\ref{CHEMEQUIL} we solve kinetic equations for all lepton and Higgs asymmetries both analytically, neglecting hypercharge fields, and numerically, including hypermagnetic fields. In Sec.~\ref{CHIRANOMPAR} we analyze the evolution of the chiral anomaly parameter $\Delta \mu (t)$ in the symmetric phase down to $t=t_\mathrm{EW}$. Our results are discussed in Sec.~\ref{DISC}.

In Appendix~\ref{ORIGIN} we interpret and compare two quantum mechanisms producing $\alpha$-helicity parameter for magnetic and hypermagnetic fields: the chiral anomaly leading to the $\alpha_\mathrm{new}$-helicity parameter in the Faraday equation \cite{Boyarsky:2011uy}, and the Chern-Simons anomaly for hypercharge fields interpreted as a polarization effect in plasma caused by the hypermagnetic field itself \cite{Semikoz:2011tm}. In Appendix~\ref{LEPTKIN} we derive kinetic equations for the lepton and Higgs asymmetries used in the main Sec.~\ref{KINETICS}. In Appendix~\ref{sphaleron} we give some formulas for the lepton number violation due to 't Hooft's anomaly in non-Abelian fields in order to explain how the sphaleron processes influence the left lepton kinetics.

\section{Equilibrium in the symmetric phase of electroweak plasma and the chiral anomaly problem\label{EQUILIBRIUM}}

The question of how large the chiral anomaly parameter $\Delta \mu (t)=\mu_{e_\mathrm{R}}(t) - \mu_{e_\mathrm{L}}(t)$ could be before EWPT is important as an input for the generation of Maxwellian magnetic fields after EWPT. In the SM plasma consisting of quarks, leptons, and one Higgs doublet $\varphi^\mathrm{T} = (\varphi^{(+)}, \varphi^{(0)} )$, with the chemical potential being in Bose distribution $\mu_0=\mu_{\varphi^{(0)}}=\mu_{\varphi^{(+)}}$,  one can expect the chemical equilibrium in symmetric phase given by the relation
\begin{equation}\label{equilibrium}
  \mu_{e_\mathrm{R}} - \mu_{e_\mathrm{L}}=-\mu_0.
\end{equation}
Equation~\eqref{equilibrium} corresponds to Higgs decays and inverse decays in reactions $e_\mathrm{L}\bar{e}_\mathrm{R}\leftrightarrow \varphi^{(0)}$ and $\nu_e^\mathrm{L}\bar{e}_\mathrm{R}\leftrightarrow \varphi^{(+)}$.
Here for the SM doublet $L_e^\mathrm{T} = (\nu_{e}^{\mathrm{L}} e_{\mathrm{L}})$ we use the equality of chemical potentials in Fermi distributions $\mu_{e_\mathrm{L}}=\mu_{\nu_e^\mathrm{L}}$.

For the case of the global equilibrium in the absence of hypermagnetic fields, similar reactions with Higgs bosons obey analogous relations both for other lepton generations, $\mu_{l_\mathrm{R}} - \mu_{l_\mathrm{L}}=- \mu_0$, $l=\mu,\tau$, and for quarks, e.g., $\mu_{u_\mathrm{L}} - \mu_{d_\mathrm{R}}=\mu_0$, $\mu_{u_\mathrm{R}} - \mu_{u_\mathrm{L}}=\mu_0$, in reactions  $q_{u_\mathrm{L}}\bar{q}_{d_\mathrm{R}}\leftrightarrow \varphi^{(+)}$ and $q_{u_\mathrm{R}}\bar{q}_{u_\mathrm{L}}\leftrightarrow \varphi^{(0)}$ correspondingly~\cite{Harvey:1990qw,Rubakov}.

We consider below only one generation with the lowest Yukawa coupling of fermions with Higgs bosons $h_e=\sqrt{2}m_e/v=2.94\times 10^{-6}$. Thus right electrons enter the equilibrium with left particles through Higgs (inverse) decays in the expanding Universe in the last instance. This is because of the high rate of chirality flip reactions $\Gamma_\mathrm{RL}\sim h_e^2T$, which becomes faster than the Hubble expansion $H\sim T^2$, $\Gamma_\mathrm{RL}>H$, at temperatures below $T_\mathrm{RL}\sim 10\ \mathrm{TeV}$.  This fact is important in scenarios where the generation of BAU proceeds through the leptogenesis, and a primordial BAU is stored in right electrons $e_\mathrm{R}$ that are being protected from washing out by sphalerons  all the way down to $T_\mathrm{RL}$. Suggesting such scenario, the authors of Ref.~\cite{Campbell:1992jd} supposed that such value $T_\mathrm{RL}$ is close to the temperature at which the sphaleron effects fall out of the equilibrium, and therefore it is possible that the $e_\mathrm{R}$ may not be transformed into $e_\mathrm{L}$ soon enough for the sphalerons to turn them into antiquarks, and thereby wipe out the remaining BAU.

In this scenario the global equilibrium~\cite{Harvey:1990qw,Rubakov} fails, and five (=5) remaining chemical potentials describe equilibrium in a hot plasma before EWPT: three $\mu_i$ for the three global charges $B/3 - L_i = \mathrm{const}$, where $i=1,2,3$ enumerates generations in SM, $\mu_\mathrm{Y}$ for the conserved hypercharge (global $\langle Y \rangle = 0$), and $\mu_{e_\mathrm{R}}$ for right electrons
$e_\mathrm{R}$ with the conservation of their lepton number $\partial_{\mu}j^{\mu}_{e_\mathrm{R}}=0$ unless $T>T_\mathrm{RL}$~\cite{Giovannini:1997eg}. Then, if one assumes the presence of large-scale hypercharge fields $Y_{\mu}$ in the symmetric phase, which are progenitors of Maxwellian fields in the broken phase, the number of right electrons is not conserved because of the Abelian anomaly~\cite{fn1}
%
\begin{equation}\label{right}
  \partial_{\mu}j^{\mu}_{e_\mathrm{R}} = \frac{g'^{2}Y_\mathrm{R}^2}{64\pi^2}Y_{\mu\nu}\tilde{Y}^{\mu\nu},
\end{equation}
where $Y_{\mu\nu}$ and $\tilde{Y}_{\mu\nu}$ are, respectively, the $U_\mathrm{Y}(1)$ hypercharge field strengths and their duals, and $Y_\mathrm{R}=-2$ is the hypercharge of the right electron.

There are no asymmetries of left leptons  and Higgs bosons  in this scenario, $\mu_{e_\mathrm{L}}=\mu_0=0$, and the chiral asymmetry~(\ref{equilibrium}) reduces to $\Delta \mu=\mu_{e_\mathrm{R}}$. For such scenario with a nonzero $e_\mathrm{R}$ asymmetry alone~\cite{Giovannini:1997eg}, sphaleron washing out BAU is absent all the way down to EWPT.

In a broadened scenario with nonzero left lepton asymmetries $\xi_{e_\mathrm{L}}=\xi_{\nu_e^\mathrm{L}}\neq 0$, where $\xi_a=\mu_a/T$, appropriate for the stage $T<T_\mathrm{RL}$~\cite{Semikoz:2011tm,Dvornikov:2011ey}, we somehow violate the equilibrium described in Ref.~\cite{Giovannini:1997eg} by five chemical potentials for five globally conserved charges. Nevertheless, it can lead only  to an additional factor of the order one $c_{\Delta}\sim 1$ that describes the dependence of $n_\mathrm{L}=(n_{e_\mathrm{L}} - n_{\bar{e}_\mathrm{L}})=\xi_{e_\mathrm{L}}T^3/6\neq 0$ on five global charges in primordial plasma. For instance, rewriting the canonical Abelian anomaly for the left doublet $L_e^\mathrm{T} = (\nu_e^\mathrm{L}, e_\mathrm{L})$,
\begin{equation}\label{left}
  \partial_{\mu}j^{\mu}_{e_\mathrm{L}} = -\frac{g'^{2}Y_\mathrm{L}^2}{64\pi^2}Y_{\mu\nu}\tilde{Y}^{\mu\nu},
  \quad
  Y_\mathrm{L}=-1,
\end{equation}
in the form $\mathrm{d}\xi_{e_\mathrm{L}}/\mathrm{d}t = -c_{\Delta}(6g'^{2}/16\pi^2T^3)(\mathbf{E}_\mathrm{Y} \cdot \mathbf{B}_\mathrm{Y})$,
we put below $c_{\Delta}=1$ simplifying the solution of our kinetic equations for the lepton and Higgs boson asymmetries.
Note that assuming a nonzero left particle asymmetry $\xi_{e_\mathrm{L}}\neq 0$, we should take into account the sphaleron processes violating lepton and baryon numbers. The competition of such processes with hypermagnetic field contribution through Abelian anomaly is one of the interesting questions touched upon in the present work.

\section{Kinetics of leptons and Higgs bosons in hypermagnetic fields\label{KINETICS}}

In Ref.~\cite{Dvornikov:2011ey} we forced the presence of zero Higgs asymmetry $n_{\varphi^{(0)}} - n_{\tilde{\varphi}^{(0)}}= T^2\mu_0/3=0$, $\mu_0=0$ considering leptogenesis with the inverse decays only, $e_\mathrm{R}\bar{e}_\mathrm{L}\to \tilde{\varphi}^{(0)}$, $e_\mathrm{R}\bar{\nu}_e^\mathrm{L}\to \varphi^{(-)}$, etc. Now let us consider both inverse Higgs decays and direct Higgs decays. The system of kinetic equations for leptons accounting for Abelian anomalies~(\ref{right}), (\ref{left}), and sphaleron processes for left leptons  takes the form
\begin{widetext}
\begin{align}\label{system}
  \frac{{\rm d}L_{e_\mathrm{R}}}{\rm dt} = &
  \frac{g'^2}{4\pi^2s} (\mathbf{E}_\mathrm{Y} \cdot \mathbf{B}_\mathrm{Y}) +
  2\Gamma_\mathrm{RL}
  \left\{
    L_{e_\mathrm{L}}-L_{e_\mathrm{R}}-\frac{[n_{\varphi^{(0)}} -
    n_{\tilde{\varphi}^{(0)}}]}{2s}
  \right\},
  \notag
  \\
  & \text{for decays (inverse decays)}
  \quad
  e_\mathrm{R}\bar{e}_\mathrm{L}\leftrightarrow \tilde{\varphi}^{(0)}
  \quad
  \text{and}
  \quad
  e_\mathrm{R}\bar{\nu}_e^\mathrm{L}\leftrightarrow \varphi^{(-)},
  \notag
  \\
  \frac{{\rm d}L_{e_\mathrm{L}}}{\rm dt} = &
  -\frac{g'^2}{16\pi^2s}(\mathbf{E}_\mathrm{Y} \cdot \mathbf{B}_\mathrm{Y}) - \frac{\Gamma_\mathrm{sph}}{2}L_{e_\mathrm{L}}+
  \Gamma_\mathrm{RL}
  \left\{
    L_{e_\mathrm{R}} - L_{e_\mathrm{L}} + \frac{[n_{\varphi^{(0)}} -
    n_{\tilde{\varphi}^{(0)}}]}{2s}
  \right\},
  \notag
  \\
  &  \text{for}
  \quad
  \bar{e}_\mathrm{R}e_\mathrm{L}\leftrightarrow \varphi^{(0)},
  \quad
  \text{as well as}
  \notag
  \\
  \frac{{\rm d}L_{\nu_e^\mathrm{L}}}{\rm dt} = &
  -\frac{g'^2}{16\pi^2s}(\mathbf{E}_\mathrm{Y} \cdot \mathbf{B}_\mathrm{Y}) -  \frac{\Gamma_\mathrm{sph}}{2}L_{\nu_e^\mathrm{L}} +
  \Gamma_\mathrm{RL}
  \left\{
    L_{e_\mathrm{R}} - L_{e_\mathrm{L}} + \frac{[n_{\varphi^{(0)}} -
    n_{\tilde{\varphi}^{(0)}}]}{2s}
  \right\},
  \notag
  \\
  & \text{for}
  \quad
  \bar{e}_\mathrm{R}\nu_e^\mathrm{L}\leftrightarrow \varphi^{(+)}.
\end{align}
\end{widetext}
Here $L_b=(n_b - n_{\bar{b}})/s$ is the lepton number, $b=e_\mathrm{R},e_\mathrm{L}, \nu_e^\mathrm{L}$, $s=2\pi^2g^*T^3/45$ is the entropy density, and $g^*=106.75$ is the number of relativistic degrees of freedom. The factor of 2 in front of the rate $\Gamma_\mathrm{RL}$ in the
first line takes into account the equivalent reaction branches.
We also included Higgs decays with the rate $\Gamma_\mathrm{D}=\Gamma_\mathrm{RL}/2$. The probability $\Gamma_\mathrm{sph} = C \alpha_\mathrm{W}^5 T$ is  given by sphaleron transitions decreasing the left lepton numbers and therefore washing out BAU, where $\alpha_\mathrm{W} = g^2/4\pi = 1/137\sin^2\theta_\mathrm{W} = 3.17 \times 10^{-2}$ is given by the gauge coupling $g = e / \sin \theta_\mathrm{W}$ in SM and $\theta_\mathrm{W}$ is the Weinberg angle. The constant $C\simeq 25$ is estimated through lattice calculations (see some comments on 't Hooft's anomaly in Appendix~\ref{sphaleron} and Chap.~11 in Ref.~\cite{Rubakov}).
Of course, for the left doublet $L_{e_\mathrm{L}}=L_{\nu_e^\mathrm{L}}$.

This system is completed by the kinetic equation for the Higgs bosons independent of Abelian anomaly inherent in fermions~\cite{fn2}
%
\begin{multline}\label{Higgs}
  \frac{\mathrm{d}}{\mathrm{d}t}[(n_{\varphi^{(0)}} - n_{\tilde{\varphi}^{(0)}})/s]
  \\
  =
  \Gamma_\mathrm{RL}
  \left\{
    L_{e_\mathrm{L}}-L_{e_\mathrm{R}}-\frac{[n_{\varphi^{(0)}} - n_{\tilde{\varphi}^{(0)}}]}{2s}
  \right\}.
\end{multline}

Note that the rate of Higgs decays (inverse decays) coincides with the rate of a lepton pair production (annihilation) having opposite sign since the creation of a pair is followed by the disappearance of a Higgs boson and vice versa.

In kinetic Eqs.~(\ref{system}) and~(\ref{Higgs}) we used the rate of all
inverse processes ~\cite{Campbell:1992jd}, which is twice bigger than for the decay ones $\Gamma_\mathrm{RL}=2\Gamma_\mathrm{D}$,
\begin{equation}\label{Gamma}
  \Gamma_\mathrm{RL} = 5.3\times 10^{-3}h_e^2
  \left(
    \frac{m_0}{T}
  \right)^2
  T =
    \frac{\Gamma_0}{2t_\mathrm{EW}}
    \frac{1 -x}{\sqrt{x}}
  .
\end{equation}
This rate vanishes just at the EWPT time $x=1$, where the variable $x=t/t_\mathrm{EW}=(T_\mathrm{EW}/T)^2$ is given by the Friedmann law. Here $h_e=2.94\times 10^{-6}$
is the Yukawa coupling for electrons, $\Gamma_0=121$, and
$m_0^2(T)=2DT^2(1-T_\mathrm{EW}^2/T^2)$ is the
temperature dependent effective Higgs mass at zero momentum and
zero Higgs vacuum expectation value. The coefficient $2D\approx 0.377$ for $m_0^2(T)$ is given by
the known masses of gauge bosons $m_\mathrm{Z}$ and $m_\mathrm{W}$, the top quark mass
$m_t$, and a still problematic zero-temperature Higgs mass, which is
estimated as $m_\mathrm{H} \sim 125\ \mathrm{GeV}$ (see Ref.~\cite{Higgsmass}). Of course, the chirality flipping rate exists after EWPT. However, that rate is due to electromagnetic processes at $T<T_\mathrm{EW}$, $\Gamma_\mathrm{em}\simeq \alpha^2_\mathrm{em}(m_e^2/3T^2)T$  when particles (electrons and positrons) acquire the nonzero mass $m_e$.

The detailed derivation of kinetic Eqs.~(\ref{system}) and~(\ref{Higgs}) accounting for chirality flip processes (without Abelian anomaly and sphaleron transitions) is given in Appendix~\ref{LEPTKIN}.

Let us rewrite Eqs.~(\ref{system}) and~(\ref{Higgs}) using the asymmetries $L_{e_\mathrm{R}}=\xi_{e_\mathrm{R}}T^3/6s$, $L_{e_\mathrm{L}}=\xi_{e_\mathrm{L}}T^3/6s$ and $(n_{\varphi^{(0)}} - n_{\tilde{\varphi}^{(0)}})/s=\xi_0T^3/3s$ as
\begin{align}\label{system2}
  \frac{\mathrm{d}\xi_{e_\mathrm{R}}}{\mathrm{d}t} = &
  \frac{3g'^2}{2\pi^2T^3}\mathbf{E}_\mathrm{Y} \cdot \mathbf{B}_\mathrm{Y} + 2\Gamma_\mathrm{RL}(-\xi_{e_\mathrm{R}} + \xi_{e_\mathrm{L}} - \xi_0),
  \nonumber\\
  \frac{\mathrm{d}\xi_{e_\mathrm{L}}}{\mathrm{d}t} = &
  -\frac{3g'^2}{8\pi^2T^3}\mathbf{E}_\mathrm{Y} \cdot \mathbf{B}_\mathrm{Y} - \frac{\Gamma_\mathrm{sph}}{2}\xi_{e_\mathrm{L}}
  \notag
  \\
  & +
  \Gamma_\mathrm{RL}
  (\xi_{e_\mathrm{R}} - \xi_{e_\mathrm{L}} + \xi_0),
  \nonumber\\
  \frac{\mathrm{d}\xi_{\nu_e^\mathrm{L}}}{\mathrm{d}t} = &
  -\frac{3g'^2}{8\pi^2T^3}\mathbf{E}_\mathrm{Y} \cdot \mathbf{B}_\mathrm{Y} - \frac{\Gamma_\mathrm{sph}}{2}\xi_{e_\mathrm{L}}
  \notag
  \\
  & +
  \Gamma_\mathrm{RL}
  (\xi_{e_\mathrm{R}} - \xi_{e_\mathrm{L}} + \xi_0),
  \nonumber\\
  \frac{\mathrm{d}\xi_{0}}{\mathrm{d}t}= &
  \Gamma_\mathrm{RL}(-\xi_{e_\mathrm{R}} + \xi_{e_\mathrm{L}} - \xi_0).
\end{align}
The third equation for neutrinos is excess since $\xi_{\nu_e^\mathrm{L}}=\xi_{e_\mathrm{L}}$. Thus, we have three equations for three chemical potentials instead of the two in Ref.~\cite{Dvornikov:2011ey}. Note that we should have $\mathrm{d}\xi_0/\mathrm{d}t<0$ for our initial conditions $\xi_{e_\mathrm{R}}(t_0)>0$ and $\xi_{e_\mathrm{L}}(t_0)=\xi_0(t_0)=0$ resulting in the {\it negative chemical potential} for the boson doublet $\varphi^\mathrm{T} = (\varphi^{(+)}, \varphi^{(0)} )$, $\mu_0<0$, as it should be.

Below we simplify the Abelian anomaly contribution $\sim (\mathbf{E}_\mathrm{Y} \cdot \mathbf{B}_\mathrm{Y})$ considering, as in Ref.~\cite{Dvornikov:2011ey}, the simplest configuration of hypermagnetic field: CS wave $Y_x=Y(t)\sin k_0z$, $Y_y=Y(t)\cos k_0z$, $Y_z=Y_0=0$. Using the generalized Ohm's law~\cite{Semikoz:2011tm}
$$
  \mathbf{E}_\mathrm{Y}=
  -{\bf V}\times \mathbf{B}_\mathrm{Y} + \eta_\mathrm{Y}\nabla\times \mathbf{B}_\mathrm{Y} -
  \alpha_\mathrm{Y}\mathbf{B}_\mathrm{Y},
$$
where $\eta_\mathrm{Y}=(\sigma_\mathrm{cond})^{-1}$ is the magnetic (hypermagnetic) diffusion coefficient, $\alpha_\mathrm{Y}$ is the hypermagnetic helicity parameter arising due to the polarization of electroweak plasma \cite{Semikoz:2011tm,Dvornikov:2011ey},
\begin{equation}\label{alpha}
  \alpha_\mathrm{Y} = \frac{g'^{2}(\mu_{e_\mathrm{R}} + \mu_{e_\mathrm{L}}/2)}{4\pi^2\sigma_\mathrm{cond}},
  \quad
  \sigma_\mathrm{cond}=100T,
\end{equation}
and we get the pseudoscalar $(\mathbf{E}_\mathrm{Y}\cdot\mathbf{B}_\mathrm{Y})$ entering the Abelian anomaly as
\begin{align}\label{product}
  (\mathbf{E}_\mathrm{Y}\cdot\mathbf{B}_\mathrm{Y}) = &
  \eta_\mathrm{Y}(\nabla\times \mathbf{B}_\mathrm{Y})\cdot \mathbf{B}_\mathrm{Y} -\alpha_\mathrm{Y} \mathbf{B}_\mathrm{Y}^2
  \notag
  \\
  & =
  \frac{B_\mathrm{Y}^2}{100}\left[\frac{k_0}{T} - \frac{g'^{2}}{4\pi^2}\left(\xi_\mathrm{R} + \frac{\xi_\mathrm{L}}{2}\right)\right].
\end{align}
Here we substituted $(\nabla\times \mathbf{B}_\mathrm{Y})\cdot \mathbf{B}_\mathrm{Y}=k_0B_\mathrm{Y}^2(t)$ for the CS wave, where $B_\mathrm{Y}(t)=k_0Y(t)$ is the hypermagnetic field amplitude.

Using the notations $y_\mathrm{R}(x)=10^4\xi_{e_\mathrm{R}}(x)$, $y_\mathrm{L}(x)=10^4\xi_{e_\mathrm{L}}(x)$, and $y_0(x)=10^4\xi_0(x)$, as well as accounting for Eq.~(\ref{product}), the system~(\ref{system2}) can be rewritten in the form that is analogous to Eq.~(3.4) in Ref.~\cite{Dvornikov:2011ey} (without contribution of neutrinos, which is identical to that of left electrons)
\begin{align}\label{system3}
  \frac{\mathrm{d}y_\mathrm{R}}{\mathrm{d}x} = &
  \left[
    B_0x^{1/2} - A_0
    \left(
      y_\mathrm{R} + \frac{y_\mathrm{L}}{2}
    \right)
  \right]
  \left(
    \frac{B_\mathrm{Y}^{(0)}}{10^{20}\ \mathrm{G}}
  \right)^2
  x^{3/2} e^{\varphi (x)}
  \notag
  \\
  & -\Gamma_0\frac{(1-x)}{\sqrt{x}}(y_\mathrm{R}-y_\mathrm{L} + y_0),
  \notag
  \\
  \frac{\mathrm{d}y_\mathrm{L}}{\mathrm{d}x} = &
  -\frac{1}{4}
  \left[
    B_0x^{1/2} - A_0
    \left(
      y_\mathrm{R} + \frac{y_\mathrm{L}}{2}
    \right)
  \right]
  \left(
    \frac{B_\mathrm{Y}^{(0)}}{10^{20}\ \mathrm{G}}
  \right)^2
  x^{3/2} e^{\varphi (x)}
  \notag
  \\
  &
  -
  \frac{5.6\times 10^7C}{\sqrt{x}}y_\mathrm{L}
  -\Gamma_0\frac{(1-x)}{2\sqrt{x}}(y_\mathrm{L}-y_\mathrm{R} - y_0),
  \notag
  \\
  \frac{\mathrm{d}y_0}{\mathrm{d}x} = &
  \frac{\Gamma_0(1-x)}{2\sqrt{x}}(y_\mathrm{L} - y_\mathrm{R} -y_0).
\end{align}
Here
\begin{equation}
  B_0=25.6
  \left(
    \frac{k_0}{10^{-7}T_\mathrm{EW}}
  \right),
  \quad
  A_0=77.6,
\end{equation}
are constants chosen for hypermagnetic fields normalized on $10^{20}\ \mathrm{G}$.

The function $e^{\varphi(x)}$ is given by the
hypermagnetic field squared
\begin{equation}
  e^{\varphi(x)} =
  \left[
    \frac{B_\mathrm{Y}(x)}{B_\mathrm{Y}^{(0)}}
  \right]^2.
\end{equation}
We also substituted the hypermagnetic field $B_\mathrm{Y}(t)=k_0Y(t)$ found as the solution of the modified Faraday equation~\cite{Semikoz:2009ye,Semikoz:2007ti} for the CS wave~\cite{fn3}
%
\begin{align}\label{Faradaysolution}
  B_\mathrm{Y}(t) = & B_\mathrm{Y}^{(0)} \exp
  \left\{
    \int_{t_0}^t[\alpha_\mathrm{Y}(t')k_0 - k_0^2\eta_\mathrm{Y}(t')]\mathrm{d}t'
  \right\} =
  \notag
  \\
  & = B_\mathrm{Y}^{(0)} \exp
  \bigg\{
    3.5
    \left(
      \frac{k_0}{10^{-7}T_\mathrm{EW}}
    \right)
    \notag
    \\
    & \times
    \int_{x_0}^x
    \bigg[
      \frac{(y_\mathrm{R} + y_\mathrm{L}/2)}{\pi}
      \notag
      \\
      & -
      0.1
      \left(
        \frac{k_0}{10^{-7}T_\mathrm{EW}}
      \right)
      \sqrt{x'}
    \bigg]
    \mathrm{d}x'
  \bigg\}.
\end{align}
Note that we do not consider here a negative value of the wave number $k_0<0$ that is allowed as well and could lead  to the lepton number violation via Abelian anomaly proportional to the pseudoscalar (\ref{product}), $(\mathbf{E}_\mathrm{Y} \cdot \mathbf{B}_\mathrm{Y}) \sim k_0^3 Y^2(t) < 0$. This is because the case $k_0<0$ corresponds to the decay of the hypermagnetic field (\ref{Faradaysolution}) instead of a real instability evolving in MHD plasma for $k_0>0$.

We choose initial conditions
at $x_0=10^{-4}$ or at $T_0=T_\mathrm{RL}$ when Higgs (inverse) decay becomes faster than the Hubble expansion $\Gamma_\mathrm{RL}>H$,
\begin{equation}
  \label{initial}y_\mathrm{R}(x_0)=10^{-6},
  \quad
  y_\mathrm{L}(x_0)=y_0(x_0)=0.
\end{equation}
Such conditions correspond to the right electron asymmetry $\xi_{e_\mathrm{R}}(x_0)=10^{-10}$ chosen at the level of baryon asymmetry.

\subsection{Conservation laws and BAU in hypermagnetic fields\label{CONSERVLAWS}}

One can see from kinetic Eq.~(\ref{system}) that in the absence of hypercharge fields the total lepton number is not conserved due to sphaleron transitions washing out the left lepton number $\mathrm{d}L_e/\mathrm{d}t=\dot{L}_{e_\mathrm{R}} + \dot{L}_{e_\mathrm{L}} + \dot{L}_{\nu_{e}^\mathrm{L}}=-\Gamma_\mathrm{sph}L_{e_\mathrm{L}}$. The baryogenesis arises through the leptogenesis due to the conservation law $B/3- L_e = \mathrm{const}$, where $B=(n_\mathrm{B}- n_{\bar{\mathrm{B}}})/s$. Accounting for Abelian anomalies in system~(\ref{system}), such baryogenesis is possible, $\dot{B}\neq 0$, since the hypermagnetic fields raise the lepton number and BAU as well  $\mathrm{d}L_e/\mathrm{d}t|_{B_\mathrm{Y}\neq 0} > 0$, $\mathrm{d}B / \mathrm{d}t|_{B_\mathrm{Y}\neq 0}>0$. This growth proceeds opposite to the competing sphaleron influence erasing $L_{e_\mathrm{L}}$ and $B$  (compare in Ref.~\cite{Dvornikov:2011ey} where we neglected sphaleron transitions).

Three global charges are conserved ($\delta_i=\mathrm{const}$),
\begin{equation}\label{laws}
  \frac{B}{3} - L_e=\delta_1,
  \quad
  \frac{B}{3} - L_{\mu}=\delta_2,
  \quad
  \frac{B}{3} - L_{\tau}=\delta_3,
\end{equation}
as well as $L_{e_\mathrm{R}}=\delta_\mathrm{R}$ well above $T_\mathrm{RL}$, $T\gg T_\mathrm{RL}$. If the initial BAU differs from zero, $B(t_0)\neq 0$, and if we assume the absence of lepton asymmetries for the second and third generations all the way down  to $T_\mathrm{EW}$, $L_{\mu}=L_{\tau}=0$,  we  find that the relation $\delta_2=\delta_3=B(x_0)/3$ is valid only for the initial time. From the first conservation law in Eq.~(\ref{laws})  one finds the change of BAU $B(t)$ at temperatures $T<T_\mathrm{RL}$. This change obeys the relations
$$
  \frac{B(t)}{3} - L_e(t)=\frac{B(t_0)}{3} - L_{e_\mathrm{R}}(t_0)=\delta_{2,3} - \delta_\mathrm{R}=\delta_1.
$$
If, for simplicity, we assume the zero initial BAU $B(t_0)=0$ or $\delta_{2,3}=0$, then finally we get the conservation law $B(t)/3-L_e(t)=-L_{e_\mathrm{R}}(t_0)$.

\begin{figure*}
  \centering
  \includegraphics[scale=1]{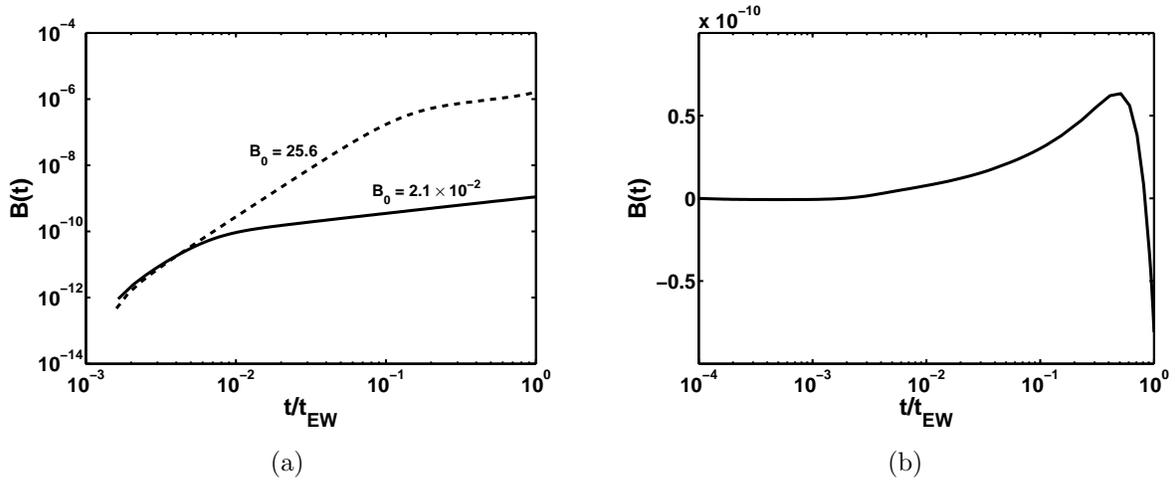}
  \caption{The baryon asymmetry $B(t)$ versus $t/t_\mathrm{EW}$ for $B_\mathrm{Y}^{(0)} = 10^{19}\ \text{G}$.
  (a) The baryon asymmetry for $B_0 = 2.1 \times 10^{-2}$ (solid line) and $B_0 = 25.6$ (dashed line).
  (b) The baryon asymmetry for $B_0 = 2 \times 10^{-3}$.}
  \label{etaBvsk}
\end{figure*}

Thus, in the present scenario, BAU sits in hypercharge fields that are decreasing due to the sphaleron processes, as follows from the sum of kinetic Eq.~(\ref{system}):
\begin{align}\label{baryon}
  B(t) = & 3\int_{t_0}^t\left[\frac{\mathrm{d}L_{e_\mathrm{R}}(t')}{\mathrm{d}t'} +
  \frac{\mathrm{d}L_{e_\mathrm{L}}(t')}{\mathrm{d}t'} +
  \frac{\mathrm{d}L_{\nu_{e\mathrm{L}}}(t')}{\mathrm{d}t'}\right]\mathrm{d}t'
  \notag
  \\
  & =
  \left(\frac{3g'^{2}}{8\pi^2}\right)\int_{t_0}^t(\mathbf{E}_\mathrm{Y}\cdot\mathbf{B}_\mathrm{Y})\frac{\mathrm{d}t'}{s}
  \notag
  \\
  & -
  3\int_{t_0}^t\Gamma_\mathrm{sph}L_{e_\mathrm{L}} \mathrm{d}t'.
\end{align}
Using the first equation in the system~(\ref{system3}), where the hypermagnetic term comes from the Abelian anomaly $\sim (\mathbf{E}_\mathrm{Y}\cdot\mathbf{B}_\mathrm{Y})$,  one obtains from Eq.~(\ref{baryon}) the baryon asymmetry in the following form:
\begin{align}\label{baryon2}
  B(x) = & 2.14\times 10^{-6}\int_{x_0}^x {\rm d}x'
  \bigg\{
    \frac{{\rm d}y_\mathrm{R} (x')}{{\rm d}x'}
    \notag
    \\
    & +
    \Gamma_0\frac{(1-x')}{\sqrt{x'}}
    \left[
      y_\mathrm{R}(x') - y_\mathrm{L}(x') + y_0(x')
    \right]
  \bigg\}
  \notag
  \\
  & -
  128C\int_{x_0}^x\frac{\mathrm{d} x'}{\sqrt{x'}}y_\mathrm{L}(x').
\end{align}
The baryon asymmetry~\eqref{baryon2} for different values of the parameter $B_0=25.6\times(k_0/10^{-7}T_\mathrm{EW})$ or for different CS wave numbers $k_0$ is shown in Fig.~\ref{etaBvsk}. Notice that for very small $k_0\ll k_\mathrm{max} = 10^{-7}T_\mathrm{EW}$, the role of the hypermagnetic field, which feeds BAU growth, becomes negligible since $B_\mathrm{Y} \sim k_0$. As a result, sphaleron transitions wash out BAU, or they diminish the Abelian anomaly leptogenesis effect in such a way that BAU can be even negative at the EWPT time $B(t_\mathrm{EW})<0$ [see curve $B(t)$ in Fig.~\ref{etaBvsk}(b) plotted for the parameter $B_0=2\times 10^{-3}$ that corresponds to $k_0=7.8\times 10^{-5}k_\mathrm{max}$].

\section{Chemical equilibrium with and without hypermagnetic fields\label{CHEMEQUIL}}

\begin{figure*}
  \centering
  \includegraphics[scale=1]{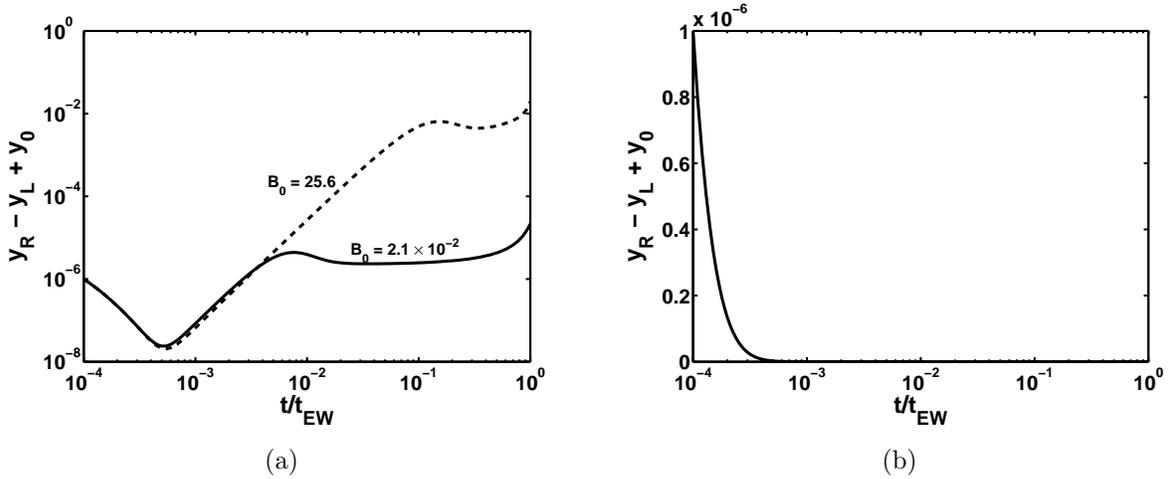}
  \caption{The function $y_\mathrm{R}-y_\mathrm{L}+y_0$ versus $t/t_\mathrm{EW}$ for lepton asymmetries $y_a=10^4(\mu_a/T)$.
  (a) The numerical solution of the system~\eqref{system3} for $B_\mathrm{Y}^{(0)} = 10^{19}\ \text{G}$. The solid line corresponds to
  $B_0 = 2.1 \times 10^{-2}$ and the dashed line to $B_0 = 25.6$.
  (b) The analytic expression for $y_\mathrm{R}-y_\mathrm{L}+y_0$ given by Eq.~\eqref{equilibrium1}.}
  \label{etaBvskrb}
\end{figure*}

Neglecting the hypermagnetic field contribution and sphaleron transitions and using the initial condition~(\ref{initial}), we easily find the solutions of kinetic equations~(\ref{system3})
\begin{eqnarray}\label{chemical}
  &&y_\mathrm{R}(x) =
  \frac{y_\mathrm{R}(x_0)}{2}[1 + e^{\Phi(x)}],
  \nonumber\\&&
  y_\mathrm{L}(x) =
  \frac{y_\mathrm{R}(x_0))}{4}[1- e^{\Phi(x)}],
  \nonumber\\&&
  y_0(x) =
  -\frac{y_\mathrm{R}(x_0))}{4}[1- e^{\Phi(x)}],
  \nonumber\\&&
  \Phi(x)=
  - 4\Gamma_0
  \left[
    (x^{1/2}-x_0^{1/2}) - \frac{(x^{3/2} - x_0^{3/2})}{3}
  \right].
\end{eqnarray}
Note that $y_0<0$, as it should be for a boson chemical potential.
Obviously the chemical equilibrium~(\ref{equilibrium}) is settled soon due to huge negative $\Phi\simeq - 4\Gamma_0= - 484$,
\begin{equation}\label{equilibrium1}
  y_\mathrm{R}-y_\mathrm{L}+y_0=y_\mathrm{R}(x_0)e^{\Phi (x)}\to 0,
\end{equation}
that happens somewhere at $x>x_\mathrm{eq} \simeq 10^{-2}$ at the temperature $T=T_\mathrm{EW}/\sqrt{x_\mathrm{eq}} \simeq 1\ \mathrm{TeV}$ before EWPT, $T>T_\mathrm{EW}$ (see right panel of Fig.~\ref{etaBvskrb}).

The numerical solution of the system~(\ref{system3}) accounting for sphalerons and in the presence of $\mathbf{B}_\mathrm{Y}\neq 0$ for the particular case of CS wave configuration is shown on the left panel in the same Fig.~\ref{etaBvskrb}.  We see that opposite to the case in Eq.~(\ref{equilibrium1}), in the presence of hypermagnetic fields and accounting for sphaleron transitions, the chemical equilibrium between leptons and Higgs's (\ref{equilibrium}) never exists: the sum $y_\mathrm{R}-y_\mathrm{L}+ y_0$ even grows in the symmetric phase when $t\to t_\mathrm{EW}$. The shorter
CS wavelength (e.g., for $B_0=25.6$ if we use the maximum wave number $k_0=10^{-7}T_\mathrm{EW}$), the larger values $(y_\mathrm{R}-y_\mathrm{L}+y_0)$ evolve up to $t_\mathrm{EW}$.

All curves in Fig.~\ref{etaBvskrb} start from the same initial condition $y_\mathrm{R}(x_0)-y_\mathrm{L}(x_0)+ y_0(x_0)=y_\mathrm{R}(x_0)= 10^{-6}$ that corresponds
to the initial right electron asymmetry $\xi_{e_\mathrm{R}}=10^{-10}$ close to the BAU value we expect at the EWPT time $x=1$.
Thus, the violation of lepton numbers in external fields through Abelian anomalies and sphaleron transitions leads to a violation of the chemical equilibrium~(\ref{equilibrium}) existing in primordial plasma when perturbative reactions (decays, scattering, etc.) are taken into account only.

\section{Chiral anomaly parameter $(y_\mathrm{R}-y_\mathrm{L})\neq 0$ in electroweak plasma before EWPT\label{CHIRANOMPAR}}

\begin{figure*}
  \centering
  \includegraphics[scale=1]{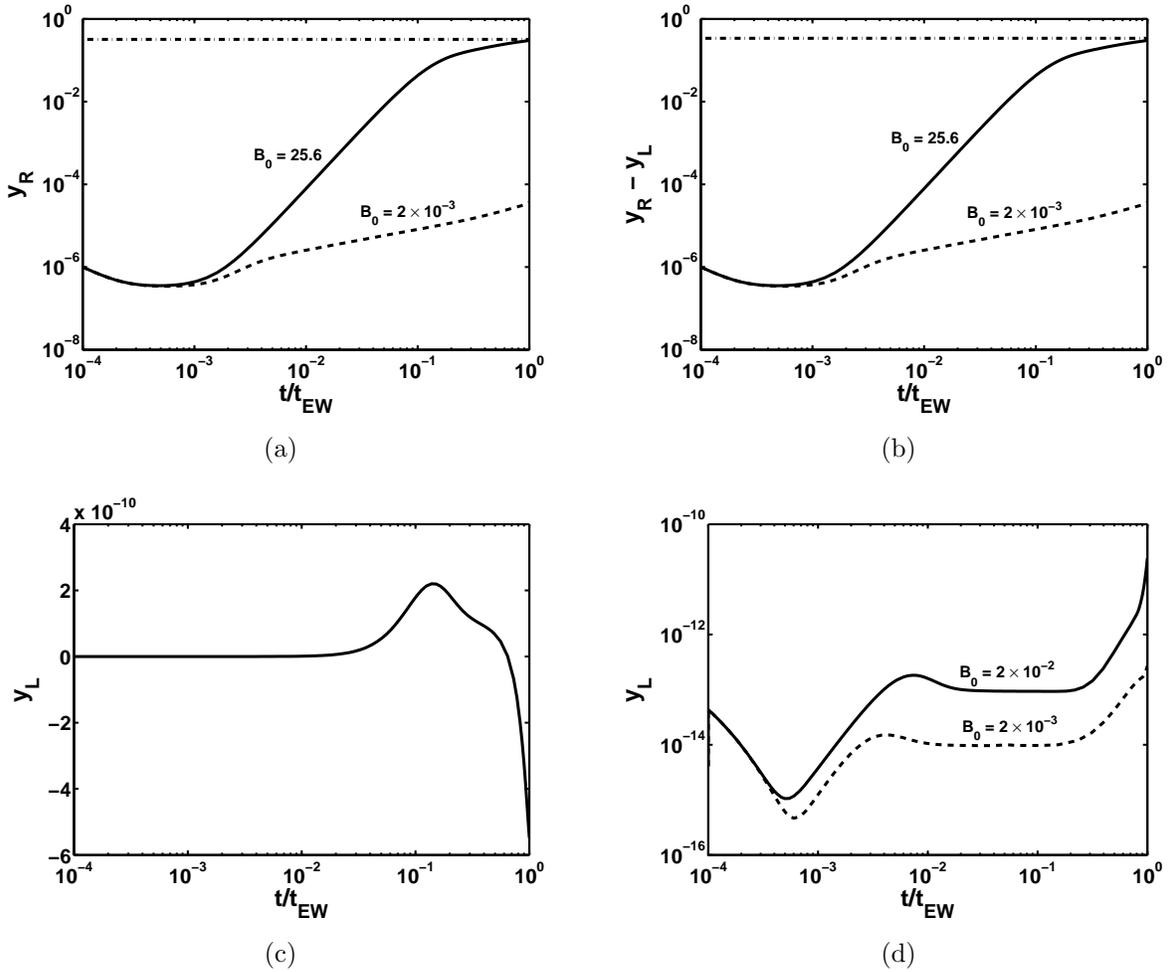}
  \caption{
  The normalized chemical potentials $y_\mathrm{R,L}$ and the chiral anomaly parameter $y_\mathrm{R}-y_\mathrm{L}$ versus $t/t_\mathrm{EW}$
  for $B_\mathrm{Y}^{(0)} = 10^{19}\ \text{G}$.
  (a) The normalized chemical potential $y_\mathrm{R}$ for $B_0 = 25.6$ (solid line) and
  $B_0 = 2 \times 10^{-3}$ (dashed line). The dash-dotted line corresponds to the asymptotic value of
  $y_\mathrm{R}=0.32$ calculated analytically in Eq.~(\ref{asymptotic}).
  (b) The chiral anomaly parameter $y_\mathrm{R}-y_\mathrm{L}$ for
  $B_0 = 25.6$ (solid line) and
  $B_0 = 2 \times 10^{-3}$ (dashed line). The dash-dotted line corresponds to the asymptotic value of
  $y_\mathrm{R} - y_\mathrm{L}$ equal to $0.34$.
  (c) The normalized chemical potential $y_\mathrm{L}$ for $B_0 = 25.6$.
  (d) The normalized chemical potential $y_\mathrm{L}$ for $B_0 = 2 \times 10^{-2}$ (solid line) and
  $B_0 = 2 \times 10^{-3}$ (dashed line).}
  \label{yRyL}
\end{figure*}

The temporal evolution of the chiral anomaly parameter $y_\mathrm{R} - y_\mathrm{L}= 10^4 (\Delta \mu/T)$ is shown in Fig.~\ref{yRyL}. There is a strong dependence on the scale $\Lambda=k_0^{-1}$ for the chosen configuration of the hypermagnetic field: the shorter CS wavelength, the bigger the chiral anomaly parameter will be. For the maximum acceptable wave number $k_0=k_\mathrm{max}=10^{-7}T_\mathrm{EW}$ ($B_0=25.6$), the chiral anomaly parameter $\Delta \mu/T$ is close to $5\times 10^{-5}$ assumed in Fig.~1 of Ref.~\cite{Boyarsky:2011uy} as a maximum initial value of the chiral anomaly parameter just after EWPT. However for longer CS wavelengths [see dashed line in Fig.~\ref{yRyL}(b) plotted for $k_0=7.8\times 10^{-5}k_\mathrm{max}$], such an initial value will occur at a negligible level ($\Delta \mu/T\sim 10^{-8}$) which can crucially change the results of Ref.~\cite{Boyarsky:2011uy}. Note that for the strongest CS field amplitude $B_\mathrm{Y}=k_\mathrm{max}Y(t)$, or in the case of the maximum Abelian anomaly leptogenesis effect,  the left lepton asymmetry $y_\mathrm{L}$ grows from the beginning due to Higgs (inverse) decays and then changes sign, $y_\mathrm{L}<0$, which is allowed for fermions, cf., Fig.~\ref{yRyL}(c).

Let us explain qualitatively the growth of the chiral anomaly shown in Figs.~\ref{yRyL}(a) and~\ref{yRyL}(b). One can simplify the kinetic equations for $\xi_{e_\mathrm{R}}$ and $\xi_{e_\mathrm{L}}$ in the system~(\ref{system2}) decoupling them. For this purpose we neglect the Higgs boson asymmetry $\xi_0=0$. We also omit the asymmetry of left leptons $\xi_{e_\mathrm{L}}=0$ in the first line of Eq.~(\ref{system2}), and the right electron in the second line of Eq.~(\ref{system2}) $\xi_{e_\mathrm{R}}=0$. For example, from the first equation in the system~(\ref{system2}), substituting the pseudoscalar value $({\bf E}_\mathrm{Y} \cdot {\bf B}_\mathrm{Y})$ for the CS wave from Eq.~(\ref{product}), one gets  the simple differential equation for the right electron asymmetry $y_\mathrm{R}=10^4\xi_{e_\mathrm{R}}$,
\begin{equation}\label{growth}
  \frac{\mathrm {d}y_\mathrm{R}}{\mathrm{d}t} + (\Gamma + \Gamma_\mathrm{B})y_\mathrm{R}= Q,
\end{equation}
where $\Gamma=2\Gamma_\mathrm{RL}$ is the chirality flip rate $\Gamma_\mathrm{B} = 6(g'^{2}/4\pi^2)^2B_\mathrm{Y}^2/100T^3$, and $Q = 6 \times 10^4 \times g'^{2}B_\mathrm{Y}^2k_0/400\pi^2T^4$ come from the second (helicity) term in Eq.~(\ref{product}) and  from the first (diffusion) term in the same Eq.~(\ref{product}). The solution of Eq.~(\ref{growth}) obtained for strong and constant hypermagnetic fields $\Gamma_\mathrm{B} \gg \Gamma$ and $\mathbf{B}_\mathrm{Y}^2 \approx \text{const}$
\begin{align}\label{growth2}
  y_\mathrm{R}(t) = &
  \left[
    y_\mathrm{R}(t_0) - \frac{Q}{\Gamma + \Gamma_\mathrm{B}}
  \right]
  e^{-(\Gamma + \Gamma_\mathrm{B})(t-t_0)}
  \notag
  \\
  & +
  \frac{Q}{\Gamma + \Gamma_\mathrm{B}}
\end{align}
gives the asymptotic growth of $y_\mathrm{R}(t)$ up to $y_\mathrm{R}(t_\mathrm{EW})$ [here for its initial value $\xi_{e_\mathrm{R}}(t_0)=0$],
\begin{align}\label{asymptotic}
  y_\mathrm{R}(t_\mathrm{EW}) = &
  \frac{Q}{\Gamma_\mathrm{B}}
  \left[
    1 - e^{-\Gamma_\mathrm{B}(t_\mathrm{EW}-t_0)}
  \right] =
  \frac{Q}{\Gamma_\mathrm{B}}
  \notag
  \\
  &
  \approx 10^{4}
  \left(
    \frac{4\pi^2}{g'^{2}}
  \right)
  \left(
    \frac{k_0}{T_\mathrm{EW}}
  \right)= 0.32.
\end{align}
Here we put $\Gamma_\mathrm{B} t_\mathrm{EW} \gg 1$ for strong fields, as well as substituted $g'^{2}=e^2/\cos^2\theta_\mathrm{W} =0.12$ and $k_0/T_\mathrm{EW}=10^{-7}$ for the case of $B_0=25.6$.

The bigger the wave number $k_0$ or the stronger the hypermagnetic field
$B_\mathrm{Y} = k_0 Y(t)$, the weaker a sphaleron influence on lepton (baryon) asymmetry is. In a wide region of wave numbers, the left lepton asymmetry $y_\mathrm{L}$ remains negligible compared with the right electron $y_\mathrm{R}$, $|y_\mathrm{L}| \ll y_\mathrm{R}$. This is due to the initial conditions~(\ref{initial}) in our scenario resulting in the right electron asymmetry $y_\mathrm{R}$ and the chiral anomaly $y_\mathrm{R} - y_\mathrm{L}$, close to what was calculated in Ref.~\cite{Dvornikov:2011ey}. However, for a long CS wave $k_0 \ll k_\mathrm{max}$, a small positive value $y_\mathrm{L} > 0$ evolving down to $T_\mathrm{EW}$ [cf., Fig.~\ref{yRyL}(d)] is sufficient to allow sphaleron transitions to wash out BAU.

\section{Discussion\label{DISC}}

In the present work we have studied how the chiral asymmetry $\Delta \mu=\mu_{e_\mathrm{R}} - \mu_{e_\mathrm{L}}\neq 0$ arises before EWPT in the scenario where, firstly, the initial right electron asymmetry $\xi_{e_\mathrm{R}}(t_0)=\mu_{e_\mathrm{R}}/T_0\simeq 10^{-10}$ provides the generation of BAU.
Secondly, if we discuss the temperatures $T\leq T_0=T_\mathrm{RL}\sim 10\ \mathrm{TeV}$, chirality flip reactions enter the equilibrium.
Then, at $t>t_0$, the violation of lepton numbers leads to a nonzero left electron asymmetry
$\xi_{e_\mathrm{L}}(t)=\mu_{e_\mathrm{L}}/T\neq 0$.
Note that we consider the lepton numbers violation due to Abelian anomalies and because of the presence of the
$\mathrm{SU}(2)_\mathrm{W}$ anomaly. The generated left electron asymmetry results in change of the primordial right electron asymmetry
$\xi_{e_\mathrm{R}}(t)$ and influences the BAU evolution.

For large scales of hypermagnetic field (for a smaller value $k_0$), sphaleron transitions are more efficient to erase BAU since the amplitude of a competitive mean hypercharge field decreases [when amplitude $B_\mathrm{Y}(t)=k_0Y(t)$ goes down], and therefore
due to Abelian anomalies, an enhancement of the lepton number ceases. Of course, the bigger the seed hypermagnetic field $B_\mathrm{Y}^{(0)}$, the bigger the lepton asymmetries and the baryon one.

While our choice of CS wave as the simplest hypermagnetic field configuration significantly simplifies the analysis of the lepton (baryon) asymmetry evolution, we should comment on some disadvantages that are appropriate to such hypercharge field.

First, a scale of hypermagnetic field for the chosen CS configuration is rather small. In order to get the BAU close to the observable value $B_\mathrm{obs}(t_\mathrm{EW})\sim 10^{-10}$ we fit the CS wave number of the order of $k_0\simeq 10^{-3}k_\mathrm{max}$, where $k_\mathrm{max}=10^{-7}T_\mathrm{EW}$ corresponds to the maximum wave number for the hypermagnetic field surviving Ohmic losses at EWPT time. The macroscopic long-range hypercharge field has the scale $\Lambda=k_0^{-1}$, which is much bigger than the mean distance between particles in plasma $T^{-1}$, and, on the other hand, is much less than the horizon size $\Lambda\ll l_H$. Here the horizon size, e.g., at the EWPT time $l_H=(M_{Pl}/1.66\sqrt{g^*})T^{-2}_\mathrm{EW}= 10^{16}/T_\mathrm{EW}$, is much bigger than the scales $k_0^{-1}=10^7/T_\mathrm{EW}$ and $k_0^{-1}\simeq (10^{10} - 10^{11})/T_\mathrm{EW}$ applied in our plots. Because of the arbitrariness of the $z$-axis direction chosen for the CS wave, such macroscopic fields are rather small-scale (random) fields and there is no anisotropy of medium. Thus, in order to get a necessary scale on the onset of galactic magnetic fields we should rely on an inverse cascade evolving after the EWPT for the Maxwellian fields~\cite{Brandenburg:1996fc} that originated from the hypercharge ones in our causal scenario. Note that such inverse cascade needs a significant amount of magnetic helicity in order to operate efficiently. For the CS wave, the helicity density
$h_\mathrm{Y} \sim k_0^3 \smallint \mathrm{d}t Y^2(t)$
decreases faster
with lowering of the wave number $k_0$ than the energy density $\rho^{(B)}_\mathrm{Y}=B_\mathrm{Y}^2/2 = k^2_0Y^2(t)/2$. In other words, the lower $k_0$, the further away the CS wave configuration is from the maximum helical field obeying the relation $k_0h_\mathrm{Y}^\mathrm{max} = 2 \rho^{(B)}_\mathrm{Y}$. This circumstance should be taken into account for a more realistic continuous hypercharge field spectrum when the conservation of the global helicity governs a spread of helicities over different scales.

In addition, our choice of CS wave does not appear to be not realistic for transition of hypercharge field to the Maxwellian one during EWPT. It was shown in Ref.~\cite{Akhmet'ev:2010ba} that
being provided by the helicity conservation only a helical field such as that given by the 3D configuration of $Y_{\mu}$ can penetrate the boundary wall separating symmetric and broken phases during the EWPT time $T\sim T_\mathrm{EW}$.
The CS wave does not penetrate such surface of a bubble of a new phase even for a strong hypermagnetic field amplitude, which, in turn, provides possibility of EWPT of the first order
for the present bounds on Higgs masses~\cite{Higgsmass}.

The evolution of the corresponding hypermagnetic helicity for an arbitrary configuration of hypermagnetic fields before EWPT $H_\mathrm{Y}=\int \mathrm{d}^3\mathbf{x} ({\bf Y}\cdot{\bf B}_\mathrm{Y})$ has been recently studied in Ref.~\cite{Semikoz:2012ka} neglecting hypermagnetic diffusion. The following magnetic helicity evolution in hot plasma at temperatures $T\ll T_\mathrm{EW}$ was analyzed in the same approximation in Ref.~\cite{Semikoz:2004rr} relying on the model proposed in Ref.~\cite{Semikoz:2003qt} for the magnetic helicity parameter $\alpha \sim G_\mathrm{F}$. The new mechanism for the $\alpha$-helicity parameter suggested in Refs.~\cite{Boyarsky:2011uy,Boyarsky:2012ex} may improve such estimates for primordial Maxwellian fields.

To resume, we estimated the chiral asymmetry $\Delta\mu=\mu_{e_\mathrm{R}}- \mu_{e_\mathrm{L}}$ arising just at EWPT $T\simeq T_\mathrm{EW}$ using the simplest configuration of the hypercharge field---the Chern-Simons wave---and taking into account both Higgs decays and inverse Higgs decays and sphaleron transitions as well. The evolution of lepton and Higgs asymmetries were studied at temperatures $T_\mathrm{EW} \leq T\leq T_\mathrm{RL}$.  The baryon and lepton asymmetries crucially depend on the CS wavelength $k_0^{-1}$.
The doubts whether it is possible to protect the baryon asymmetry of the Universe in the symmetric phase by temporarily storing BAU in the asymmetry of the $e_\mathrm{R}$ species are dispelled in the case of strong hypermagnetic fields.
The washing out of BAU by the sphaleron transitions due to the involvement of left particles at $T<T_\mathrm{RL}$ through Higgs (inverse) decays is not dangerous in a wide region of CS wave numbers. A strong seed hypermagnetic field $B_\mathrm{Y}^{(0)}$ is needed to support all such issues while dynamo amplification turns out rather negligible for the CS wave. The amplification of hypermagnetic fields by mechanisms beyond the SM, assuming, e.g., a new pseudoscalar field coupled to hypercharge topological number density~\cite{Brustein:1999rk}, is not considered here. Within the framework of the SM model, the 3D configuration of a hypermagnetic field seems to be much more productive for problems under consideration that remain as the challenge for us in the future.

\begin{acknowledgments}
We acknowledge Jose Valle, Dmitry Sokoloff, and Valery Rubakov for fruitful discussions. M.D. is thankful to FAPESP (Brazil) for a grant.
\end{acknowledgments}

\appendix

\section{The origin of the chiral anomaly for a Maxwellian field and the CS anomaly for a hypermagnetic field\label{ORIGIN}}

The chiral anomaly parameter $\Delta \mu=(\mu_{e_\mathrm{R}} - \mu_{e_\mathrm{L}})\neq 0$ leads to an additional contribution to the current in the Maxwell equation
\begin{equation}\label{1stMaxeq}
  -\frac{\partial \mathbf{E}}{\partial t} + \nabla\times \mathbf{B} =
  \sigma_\mathrm{cond} \mathbf{E} + \frac{\alpha_\mathrm{em}}{\pi}\Delta \mu \mathbf{B},  
\end{equation}
where the last {\it pseudovector} current $\mathbf{j}=(\alpha_\mathrm{em}\Delta \mu /\pi)\mathbf{B}$ is coming, e.g., from the energy balance under chirality flip for {\it massless} particles \cite{Nielsen,Kharzeev}
\begin{equation}\label{enbal}
  \int \mathrm{d}^3 \mathbf{x} (\mathbf{j}\cdot\mathbf{E}) =
  \frac{\alpha_\mathrm{em} \Delta \mu}{\pi} \int \mathrm{d}^3 \mathbf{x}
  (\mathbf{E}\cdot\mathbf{B}). 
\end{equation}
Note that $\alpha_\mathrm{em}$ and $\sigma_\mathrm{cond}$ were defined in Sec.~\ref{INTR}.

Using the Bianchi identity $\partial_t \mathbf{B}= -\nabla\times \mathbf{E}$, one finds from Eq.~\eqref{1stMaxeq}, neglecting in the MHD approach the displacement current $\partial_t\mathbf{E}=0$,  the modified Faraday equation~\cite{Boyarsky:2011uy}
\begin{equation}
  \frac{\partial \mathbf{B}}{\partial t} =
  \left(
    \frac{\alpha_\mathrm{em} \Delta \mu}{\pi\sigma_\mathrm{cond}}
  \right)
  \nabla\times \mathbf{B} + \frac{1}{\sigma_\mathrm{cond}}\nabla^2\mathbf{B}, 
\end{equation}
which governs the evolution of the magnetic field after EWPT at temperatures $10\ \mathrm{MeV} < T < T_\mathrm{EW}$.

Let us comment on the energy balance Eq.~\eqref{enbal}. Note that in the rhs of Eq.~\eqref{enbal} the Adler anomaly for the right-handed electrons (positrons)
\begin{equation*}
  \frac{\partial j_\mathrm{R}^{\mu}}{\partial x^{\mu}} =
  + \frac{e^2}{16\pi^2}F_{\mu\nu}\tilde{F}^{\mu\nu} =
  \frac{\alpha_\mathrm{em}}{\pi}(\mathbf{E}\cdot\mathbf{B})
\end{equation*}
and for the left-handed ones~\cite{fn4}
%
\begin{equation*}
  \frac{\partial j_\mathrm{L}^{\mu}}{\partial x^{\mu}} =
  - \frac{e^2}{16\pi^2}F_{\mu\nu}\tilde{F}^{\mu\nu} =
  -\frac{\alpha_\mathrm{em}}{\pi}(\mathbf{E}\cdot\mathbf{B})
\end{equation*}
being combined for the pseudovector $j^{\mu}_\mathrm{R} - j^{\mu}_\mathrm{L} = \langle \bar{\Psi}\gamma^{\mu}\gamma^5\Psi \rangle$ in uniform medium  as
\begin{equation}\label{nLRasym}
  \frac{{\rm d}}{{\rm d}t}(n_\mathrm{R} - n_\mathrm{L}) =
  \left(\frac{2\alpha_\mathrm{em}}{\pi}\right)(\mathbf{E}\cdot\mathbf{B}) 
\end{equation}
defines the rate of the chirality flip per unit time per volume and the corresponding energy cost $(\Delta \mu/2) \mathrm{d}(n_\mathrm{R}-n_\mathrm{L})/\mathrm{d}t= (\mathbf{j}\cdot\mathbf{E})$ for such effect. Hence, multiplying by $\Delta \mu/2$ and integrating Eq.~\eqref{nLRasym} over volume, one derives the energy balance~\eqref{enbal}. While separating $\mathbf{E}$ in both sides in the integrand of~\eqref{enbal}, one obtains the pseudovector current $\mathbf{j}$ in the Maxwell Eq.~\eqref{1stMaxeq}.

The CS anomaly term in the SM Lagrangian for the hypercharge field $Y_{\mu}$ also leads to the pseudovector contribution in the Maxwell equation~\cite{Dvornikov:2011ey} (here at temperatures $T_\mathrm{EW} < T < T_\mathrm{RL}$):
\begin{align}\label{1stMaxeqmod}
  -\frac{\partial \mathbf{E}_\mathrm{Y}}{\partial t} + \nabla\times \mathbf{B}_\mathrm{Y} = &
  \sigma_\mathrm{cond} \mathbf{E}_\mathrm{Y} + \frac{\alpha'}{\pi}
  \left(
    \mu_{e_\mathrm{R}} +\frac{\mu_{e_\mathrm{L}}}{2}
  \right)
  \mathbf{B}_\mathrm{Y},
  \notag
  \\
  \alpha' = & \frac{g'^{2}}{4\pi}, 
\end{align}
where the last term has {\it polarization origin} in the presence of hypermagnetic fields in primordial plasma~\cite{Semikoz:2011tm}. Notice that the Adler anomaly used in Eq.~\eqref{enbal} corresponds to the difference of mean densities of the right- and left-handed fermions that is {\it pseudoscalar}~\cite{fn5}
$n_\mathrm{R} - n_\mathrm{L} = \langle \Psi^{+{}}_\mathrm{R} \Psi_\mathrm{R} \rangle - \langle \Psi^{+{}}_\mathrm{L} \Psi_\mathrm{L} \rangle = \langle \Psi^{+{}} \gamma^5 \Psi \rangle$ with $\Psi_\mathrm{R,L} = (1 \pm \gamma^5) \Psi / 2$, while the CS term in Eq.~(\ref{1stMaxeqmod}) is given by the {\it mean spin}
$\langle \Psi^{+{}} \bm{\alpha} \gamma^5 \Psi \rangle = \langle \Psi^{+{}} \bm{\Sigma} \Psi \rangle = \mathbf{M}/\mu_\mathrm{B}$, where {\it pseudovector} $\mathbf{M} \sim \mathbf{B}_\mathrm{Y}$ is the magnetization of medium in a hypermagnetic field $\mathbf{B}_\mathrm{Y}$.

Let us stress the similarity of both chiral magnetic effects leading to anomalous terms in Eqs.~(\ref{1stMaxeq}) and~(\ref{1stMaxeqmod}). They are caused by a polarization mechanism provided by the main Landau level contribution to the additional current in the Maxwell equation (compare in Ref.~\cite{Vilenkin}).

\section{Lepton kinetics with Higgs bosons in the absence of hypercharge fields\label{LEPTKIN}}

In this appendix we briefly discuss the lepton kinetics in the presence of Higgs bosons without hypermagnetic fields and Abelian anomaly and omitting sphaleron terms. In particular,  we explain in detail the derivation of Eqs.~(\ref{system}) and~(\ref{Higgs}).

\paragraph{Right electrons}

The reactions contributing to the right electrons dynamics are
(i) the inverse decays $e_\mathrm{R}\bar{e}_\mathrm{L}\to \tilde{\varphi}^{(0)}$ and $e_\mathrm{R}\bar{\nu}_e^\mathrm{L}\to \varphi^{(-)}$
and
(ii) the direct decays $\tilde{\varphi}^{(0)}\to e_\mathrm{R}\bar{e}_\mathrm{L}$ and $\varphi^{(-)}\to e_\mathrm{R}\bar{\nu}_e^\mathrm{L}$. We remind that when particles annihilate, a ``minus'' sign appears in front of the $\Gamma_\mathrm{RL}$ term. Taking into account that $n_{\tilde{\varphi}^{(0)}}=n_{\varphi^{(-)}}$, as well as the equivalence $n_{\bar{e}_\mathrm{L}}=n_{\bar{\nu}_e^\mathrm{L}}$, one gets a factor of 2 in the kinetic equation accounting for the two channels:
\begin{equation}\label{elrkineq}
  \frac{\mathrm{d}}{\mathrm{d}t}
  \left(
    \frac{n_{e_\mathrm{R}}}{s}
  \right)=
  2\Gamma_\mathrm{RL}[-n_{e_\mathrm{R}} - n_{\bar{e}_\mathrm{L}}]/s +
  2\Gamma_\mathrm{D}\frac{n_{\tilde{\varphi}^{(0)}}}{s}. 
\end{equation}
Here the Bose distribution for the Higgs doublet $\varphi^\mathrm{T} = (\varphi^{(+)}, \varphi^{(0)})$  is given by the chemical potential $\mu_0=  \mu_{+} =  - \mu_{-} = -\mu_{\tilde{(0)}}$ or $\mu_0=\mu_+$ for $\varphi$, while for the antiparticle $\tilde{\varphi}^\mathrm{T} = (\varphi^{(-)}, \tilde{\varphi}^{(0)})$
one gets $\mu_-=\mu_{\tilde{(0)}}=-\mu_0$. Note that equilibrium, e.g., in the reaction $e_\mathrm{R}\bar{e}_\mathrm{L}\leftrightarrow \tilde{\varphi}^{(0)}$,
would correspond to the right relation of chemical potentials neglecting hypermagnetic fields as given by Eq.~(\ref{equilibrium}) , $\mu_{e_\mathrm{R}} - \mu_{e_\mathrm{L}}=-\mu_0$.

\paragraph{Right positrons}

The reactions contributing to the right positrons dynamics are
(i) the inverse decays $\bar{e}_\mathrm{R}e_\mathrm{L}\to \varphi^{(0)}$ and $\bar{e}_\mathrm{R}\nu_e^\mathrm{L}\to \varphi^{(+)}$, as well as
(ii) the decays $\varphi^{(0)}\to \bar{e}_\mathrm{R} e_\mathrm{L}$ and $\varphi^{(+)}\to \bar{e}_\mathrm{R}\nu_e^\mathrm{L}$. Analogously to the right electrons case, we take into account that $n_{\varphi^{(0)}}=n_{\varphi^{(+)}}$ and $n_{e_\mathrm{L}}=n_{\nu_e^\mathrm{L}}$. Finally, we obtain the following kinetic equation accounting for the two channels:
\begin{equation}\label{posrkineq}
  \frac{\mathrm{d}}{\mathrm{d}t}
  \left(
    \frac{n_{\bar{e}_\mathrm{R}}}{s}
  \right)=
  2\Gamma_\mathrm{RL}[-n_{\bar{e}_\mathrm{R}} - n_{e_\mathrm{L}}]/s +
  2\Gamma_\mathrm{D}\frac{n_{\varphi^{(0)}}}{s}. 
\end{equation}
Subtracting Eq.~\eqref{posrkineq} from Eq.~\eqref{elrkineq} and taking into account that $L_{e_\mathrm{R}}=[n_{e_\mathrm{R}} - n_{\bar{e}_\mathrm{R}}]/s$ and $L_{e_\mathrm{L}}=[n_{e_\mathrm{L}} - n_{\bar{e}_\mathrm{L}}]/s$, one gets the equation similar to that derived in Ref.~\cite{Campbell:1992jd} and in our previous work~\cite{Dvornikov:2011ey} if we omit the Abelian anomaly [see the first line in Eq.~(\ref{system}) in Sec.~\ref{KINETICS}],
\begin{equation}
  \frac{\mathrm{d}L_{e_\mathrm{R}}}{\mathrm{d}t} =
  2\Gamma_\mathrm{RL}(L_{e_\mathrm{L}} - L_{e_\mathrm{R}}) +
  2\Gamma_\mathrm{D} [n_{\bar{\varphi}^{(0)}} - n_{\varphi^{(0)}}]/s. 
\end{equation}
In equilibrium, $\mathrm{d}L_{e_\mathrm{R}}/\mathrm{d}t=0$ accounting for $\Gamma_\mathrm{RL}=2\Gamma_\mathrm{D}$, we get the correct relation~(\ref{equilibrium}).

\paragraph{Left electrons}

We should take into account (i) the inverse decay $\bar{e}_\mathrm{R}e_\mathrm{L}\to \varphi^{(0)}$ and (ii) the decay, $\varphi^{(0)}\to \bar{e}_\mathrm{R}e_\mathrm{L}$, which give
\begin{equation}\label{ellkineq}
  \frac{\mathrm{d}}{\mathrm{d}t}
  \left(
    \frac{n_{e_\mathrm{L}}}{s}
  \right)=
  \Gamma_\mathrm{RL}[- n_{\bar{e}_\mathrm{R}} - n_{e_\mathrm{L}}]/s + \Gamma_\mathrm{D}\frac{n_{\varphi^{(0)}}}{s}. 
\end{equation}

\paragraph{Left positrons}

In this case (i) the inverse decay $\bar{e}_\mathrm{L}e_\mathrm{R}\to \tilde{\varphi}^{(0)}$ and (ii) the decay $\tilde{\varphi}^{(0)}\to \bar{e}_\mathrm{L}e_\mathrm{R}$ give the following contributions:
\begin{equation}\label{poslkineq}
  \frac{\mathrm{d}}{\mathrm{d}t}
  \left(
    \frac{n_{\bar{e}_\mathrm{L}}}{s}
  \right)=
  \Gamma_\mathrm{RL}(- n_{e_\mathrm{R}} - n_{\bar{e}_\mathrm{L}})/s + \Gamma_\mathrm{D}\frac{n_{\tilde{\varphi}^{(0)}}}{s}. 
\end{equation}
Subtracting Eq.~\eqref{poslkineq} from Eq.~\eqref{ellkineq} one gets
\begin{equation}
  \frac{\mathrm{d}L_{e_\mathrm{L}}}{\mathrm{d}t} =
  \Gamma_\mathrm{RL}(L_{e_\mathrm{R}} - L_{e_\mathrm{L}}) +
  \Gamma_\mathrm{D}
  [n_{\varphi^{(0)}}- n_{\tilde{\varphi}^{(0)}}]/s. 
\end{equation}
In the equilibrium  $\mathrm{d}L_{e_\mathrm{L}}/\mathrm{d}t=0$ accounting for $\Gamma_\mathrm{RL}=2\Gamma_\mathrm{D}$, we get the correct relation for chemical potentials $\mu_{e_\mathrm{R}} - \mu_{e_\mathrm{L}} + \mu_0=0$.

Let us derive the Higgs boson kinetic equations. In  Eq.~\eqref{ellkineq}, the boson $\varphi^{(0)}$ decays into the pair $\bar{e}_\mathrm{R}e_\mathrm{L}$. Hence, the boson vanishes, which increases the population of $e_\mathrm{L}$ and $\bar{e}_\mathrm{R}$ (in kinetics of left electron $e_\mathrm{L}$). In the kinetic equation for $\varphi^{(0)}$ itself, such term enters with the opposite sign as $- \Gamma_\mathrm{D}n_{\varphi^{(0)}}$ (boson disappears). Analogously, the inverse decay term should have the different sign in boson kinetics: it becomes $+ \Gamma_\mathrm{RL}(n_{e_\mathrm{R}} + n_{e_\mathrm{L}})/s$ since the pair $\bar{e}_\mathrm{R}e_\mathrm{L}$ annihilates into $\varphi^{(0)}$ increasing the population of the neutral bosons $\varphi^{(0)}$. Therefore, one obtains from Eq.~\eqref{ellkineq},
\begin{equation}\label{phi0kineq}
  \frac{\mathrm{d}}{\mathrm{d}t}
  \left(
    \frac{n_{\varphi^{(0)}}}{s}
  \right)=
  \Gamma_\mathrm{RL}[n_{\bar{e}_\mathrm{R}} + n_{e_\mathrm{L}}]/s +
  \Gamma_\mathrm{D}
  \left(
    -\frac{n_{\varphi^{(0)}}}{s}
  \right). 
\end{equation}
Analogously from Eq.~\eqref{poslkineq}, changing sign on the right-hand side, one obtains the kinetic equation for the Higgs antiparticle $\tilde{\varphi}^{(0)}$:
\begin{equation}\label{antiphi0kineq}
  \frac{\mathrm{d}}{\mathrm{d}t}
  \left(
    \frac{n_{\tilde{\varphi}^{(0)}}}{s}
  \right)=
  \Gamma_\mathrm{RL}(n_{e_\mathrm{R}} +
  n_{\bar{e}_\mathrm{L}})/s + \Gamma_\mathrm{D}
  \left(
    -\frac{n_{\tilde{\varphi}^{(0)}}}{s}
  \right). 
\end{equation}
Subtracting Eq.~\eqref{antiphi0kineq} from Eq.~\eqref{phi0kineq} and accounting for $\Gamma_\mathrm{D}=\Gamma_\mathrm{RL}/2$, we derive the kinetic equation for the Higgs boson asymmetry [see Eq.~(\ref{Higgs}) in Sec.~\ref{KINETICS}]:
\begin{multline}\label{totphi0kineq}
  \frac{\mathrm{d}}{\mathrm{d}t}[(n_{\varphi^{(0)}} - n_{\tilde{\varphi}^{(0)}})/s]
  \\
  =
  \Gamma_\mathrm{RL}
  \left\{
    L_{e_\mathrm{L}}-L_{e_\mathrm{R}}-\frac{[n_{\varphi^{(0)}} - n_{\tilde{\varphi}^{(0)}}]}{2s}
  \right\}. 
\end{multline}

\paragraph{Left neutrinos}

For left neutrinos we account for (i) the inverse decay $\bar{e}_\mathrm{R}\nu_e^\mathrm{L}\to \varphi^{(+)}$ and (ii) the decay $\varphi^{(+)}\to \bar{e}_\mathrm{R}\nu_e^\mathrm{L}$, which
give
\begin{equation}\label{nulkineq}
  \frac{\mathrm{d}}{\mathrm{d}t}
  \left(
    \frac{n_{\nu_e^\mathrm{L}}}{s}
  \right)=
  \Gamma_\mathrm{RL}[- n_{\bar{e}_\mathrm{R}} - n_{\nu_e^\mathrm{L}}]/s + \Gamma_\mathrm{D}\frac{n_{\varphi^{(+)}}}{s}. 
\end{equation}

\paragraph{Left antineutrinos}

In this case (i) the inverse decay $e_\mathrm{R}\bar{\nu}_e^\mathrm{L}\to \varphi^{(-)}$ and (ii) the decay $\varphi^{(-)}\to e_\mathrm{R}\bar{\nu}_e^\mathrm{L}$
contribute to the kinetic equation as
\begin{equation}\label{antinulkineq}
  \frac{\mathrm{d}}{\mathrm{d}t}
  \left(
    \frac{n_{\bar{\nu}_e^\mathrm{L}}}{s}
  \right)=
  \Gamma_\mathrm{RL}[-n_{e_\mathrm{R}} - n_{\bar{\nu}_e^\mathrm{L}}]/s + \Gamma_\mathrm{D}\frac{n_{\varphi^{(-)}}}{s}. 
\end{equation}
Subtracting Eq.~\eqref{antinulkineq} from Eq.~\eqref{nulkineq}, one gets
\begin{equation}\label{totnukineq}
  \frac{\mathrm{d}L_{\nu_e^\mathrm{L}}}{\mathrm{d}t} =
  \Gamma_\mathrm{RL}(L_{e_\mathrm{R}} - L_{e_\mathrm{L}}) +
  \Gamma_\mathrm{D}
  [n_{\varphi^{(+)}}- n_{\varphi^{(-)}}]/s, 
\end{equation}
where we took into account that $L_{\nu_e^\mathrm{L}}= L_{e_\mathrm{L}}$ or $n_{\nu_e^\mathrm{L}}=n_{e_\mathrm{L}}$. Of course, $n_{\varphi^{(+)}}=n_{\varphi^{(0)}}$ with the chemical potential $\mu_0$ in Bose distribution for the doublet
$\varphi^\mathrm{T} = ( \varphi^{(+)}, \varphi^{(0)})$, and  $n_{\varphi^{(-)}}=n_{\tilde{\varphi}^{(0)}}$ with the chemical potential $-\mu_0$ for the c.c. doublet $\tilde{\varphi}^\mathrm{T} = (\varphi^{(-)}, \tilde{\varphi}^{(0)})$.

For charged Higgs, which are described by Eqs.~\eqref{nulkineq}-\eqref{totnukineq}, using arguments like in the derivation of Eq.~\eqref{totphi0kineq} we obtain
from Eqs.~\eqref{nulkineq} and~\eqref{antinulkineq} the kinetic equation which is identical to Eq.~\eqref{totphi0kineq}, since $n_{\varphi^{(+)}}=n_{\varphi^{(0)}}$ and $n_{\varphi^{(-)}}=n_{\tilde{\varphi}^{(0)}}$,
\begin{multline}
  \frac{\mathrm{d}}{\mathrm{d}t}
  [(n_{\varphi^{(+)}} - n_{\varphi^{(-)}})/s]
  \\
  =
  \Gamma_\mathrm{RL}
  \left\{
    L_{e_\mathrm{L}}-L_{e_\mathrm{R}}-\frac{[n_{\varphi^{(+)}} - n_{\varphi^{(-)}}]}{2s}
  \right\}. 
\end{multline}

\section{$\mathrm{SU}(2)_\mathrm{W}$ anomaly and left fermion number violation\label{sphaleron}}

Let us use Eq.~(12-174a) in Ref.~\cite{Zuber} written for the pseudovector current of the one generation of massless fermions,
$j^{\mu}_5 = j^{\mu}_\mathrm{R} - j^{\mu}_\mathrm{L} = \bar{\psi}\gamma^{\mu}\gamma^5\psi$,
\begin{equation}\label{pseudovector}
  \partial_{\mu} j^{\mu}_5 = \partial_{\mu}
  [j^{\mu}_\mathrm{R} - j^{\mu}_\mathrm{L}] = - \frac{g^2}{16\pi^2}F^{\mu\nu}_a\tilde{F}_{\mu\nu a}.
\end{equation}
Here $j^{\mu}_\mathrm{R}=\bar{\psi}\gamma^{\mu}(1 + \gamma^5)\psi/2$ and $j^{\mu}_\mathrm{L}=\bar{\psi}\gamma^{\mu}(1 - \gamma^5)\psi/2$ are the right and left fermion currents correspondingly~\cite{fn6}.
Adding the equality~(\ref{pseudovector}) with the anomaly for leptons of the first generation given by Eq.~(11.12) in Ref.~\cite{Rubakov}
(see also~\cite{fn7}),
%
\begin{equation}\label{violation}
  \partial_{\mu} j^{\mu}_{\mathrm{L}_e} =
  \partial_{\mu}[j^{\mu}_\mathrm{R} + j^{\mu}_\mathrm{L} ] =
  \frac{g^2}{16\pi^2}F^{\mu\nu}_a\tilde{F}_{\mu\nu a},
\end{equation}
one gets the well-known issue  $\partial_{\mu}j^{\mu}_\mathrm{R}=0$ that guarantees the conservation of the right electron current in the absence of hypermagnetic fields. On the other hand, subtracting Eq.~(\ref{pseudovector}) from Eq.~(\ref{violation}) we get the violation of the left lepton current in the same non-Abelian fields,
\begin{equation}\label{leftcurrent}
  \partial_{\mu}j^{\mu}_\mathrm{L} =
  \partial_{\mu}[j^{\mu}_{e_\mathrm{L}} + j^{\mu}_{\nu_{e}^\mathrm{L}}] =
  \frac{g^2}{16\pi^2}F^{\mu\nu}_a\tilde{F}_{\mu\nu a}.
\end{equation}
Notice that here neutrino and electron currents are equivalent $j^{\mu}_{e_\mathrm{L}} = j^{\mu}_{\nu_{e}^\mathrm{L}}=j^{\mu}_\mathrm{L}/2$, as  seen from the following representation for the left field doublet $\hat{L}^\mathrm{T}=(\hat{\nu}_{e}^\mathrm{L}, \hat{e}_\mathrm{L})$,
\[
  \hat{L}=\frac{1}{\sqrt{a^2 + b^2}}
  \left(
    \begin{array}{c}
      a\\b
    \end{array}
  \right)
  \hat{\psi}_\mathrm{L},
\]
where we may put $a=-b=1$ for the isospin column. Then, using the standard field operator in the Schr\"{o}dinger representation
\begin{align*}
  \hat{\psi}_\mathrm{L}(\mathbf{x}) = & \frac{1}{(2\pi)^{3/2}}
  \sum_{r}
  \int
  \frac{\mathrm{d}^3\mathbf{p}}{\sqrt{2\varepsilon_p}}
  [\hat{b}_r(\mathbf{p}) u_\mathrm{L}^r(\mathbf{p}) e^{\mathrm{i}\mathbf{px}}
  \\
  & +
  \hat{d}_r^\dag(\mathbf{p}) v_\mathrm{L}^r(\mathbf{p}) e^{-\mathrm{i}\mathbf{px}}],
\end{align*}
one finds the current asymmetry $j^{\mu}_\mathrm{L}(\mathbf{x},t) = \mathrm{Tr} [\hat{\rho}(t)\hat{\bar{\psi}}_\mathrm{L} \gamma^{\mu}\hat{\psi}_\mathrm{L}] = j^{\mu}_{l_\mathrm{L}} - j^{\mu}_{\bar{l}_\mathrm{L}}$, where $\hat{\rho}(t)$ is the nonequilibrium statistical operator obeying the Liouville equation. Here the currents
\begin{equation*}
  j^{\mu}_{l_\mathrm{L},\bar{l}_\mathrm{L}}(\mathbf{x},t) =
  \int \frac{\mathrm{d}^3\mathbf{p}}{(2\pi)^3}
  \frac{p^{\mu}}{\varepsilon_p} f^{(l_\mathrm{L},\bar{l}_\mathrm{L})}(\mathbf{p},\mathbf{x},t),
\end{equation*}
are given by the Wigner distribution functions
\begin{equation*}
  f^{(l_\mathrm{L},\bar{l}_\mathrm{L})}(\mathbf{p},\mathbf{x},t) =
  \mathrm{Tr}
  \left[
    \sum_{\mathbf{k}}e^{\mathrm{i}\mathbf{kx}}
    f^{(l_\mathrm{L},\bar{l}_\mathrm{L})}_{\mathbf{p}+\mathbf{k}/2,r';\mathbf{p}-\mathbf{k}/2,r}(t)
  \right],
\end{equation*}
which, in turn, are given by the distribution functions in the momentum representation
$f^{(l_\mathrm{L})}_{\mathbf{p}'r';\mathbf{p}r}(t) = \mathrm{Tr} \left[ \hat{\rho}(t)\hat{b}^\dag_r(\mathbf{p}) \hat{b}_{r'}(\mathbf{p}') \right]$ for particles and  $f^{(\bar{l}_\mathrm{L})}_{\mathbf{p}' r'; \mathbf{p}r}(t) = \mathrm{Tr} \left[ \hat{\rho}(t) \hat{d}^\dag_r(\mathbf{p}) \hat{d}_{r'}(\mathbf{p}') \right]$ for antiparticles.

The violation of the left lepton number $L_{e_\mathrm{L}}$ in non-Abelian fields due to the $\mathrm{SU}(2)_\mathrm{W}$ anomaly~(\ref{leftcurrent}) proceeds with the sphaleron transition probability $\Gamma_\mathrm{sph}$ as we used in kinetic equations~(\ref{system}).


\begin{thebibliography}{50}
\bibitem{fn2}
  Note that we took into account the expansion of the Universe, since for any kind of asymmetries normalized on
  the entropy including lepton numbers, the  following relations hold:
  %
  \begin{align*}
    s\frac{\mathrm{d}}{\mathrm{d}t}
    \left(
      \frac{n_b - n_{\bar{b}}}{s}
    \right) & =
    \frac{\partial (n_b - n_{\bar{b}})}{\partial t} + 3H(n_b - n_{\bar{b}}),
    \\
    H=\frac{\dot{a}}{a}= - \frac{\dot T}{T}, &
    \quad
    \text{and}
    \quad
    \frac{\dot{s}}{s}= -3H,
  \end{align*}
  %
  where $n_b$ is an arbitrary density.
\bibitem{fn3}
  In numerical estimates, we substitute either the parameter
  $k_0/(10^{-7}\ T_\mathrm{EW}) = 1$ that is the upper limit for the CS wave
  number $k_0\leq 10^{-7}T_\mathrm{EW}$ to avoid Ohmic dissipation of
  hypermagnetic field, or $k_0/(10^{-7}\ T_\mathrm{EW}) \simeq (10^{-3}-10^{-4})$ to get observable baryon asymmetry
  $B=0.87\times 10^{-10}$ at the EWPT time $x=1$. Dynamo amplification is negligible in both cases.
\bibitem{fn5}
  With respect to the Lorentz transformation, this difference is the time component of the four-pseudovector averaged over the Fermi distribution,
  $\langle\bar{\psi}\gamma^0\gamma^5\psi\rangle$, while the true pseudoscalar should have the form $\bar{\psi}\gamma^5\psi$.
  If we average the pseudoscalar term over Fermi distributions in a uniform medium, e.g., in uniform plasma with magnetic field neglecting plasma dispersion, it vanishes,
  $\langle\bar{\psi}\gamma^5\psi\rangle=0$. While in an inhomogeneous medium, one gets the nonzero result at the perturbative level
  (e.g., for magnons in a ferromagnet), $\langle\bar{\psi}\gamma^5\psi\rangle_t = - \mathrm{i}\nabla\cdot\mathbf{M}(\mathbf{x}, t)/e\neq 0$,
  where $\mathbf{M}(\mathbf{x}, t)$ is the magnetization.
\bibitem{fn6}
  We use the same notations that coincide with Refs.~\cite{Zee,Zuber}.
\bibitem{fn7}
  Note that anomaly~(\ref{violation}) does not depend on $\gamma^5$ at all.
\end{thebibliography}
\end{document}